# MICROWAVE SINTERING OF COMPLEX SHAPES: FROM MULTIPHYSICS SIMULATION TO IMPROVEMENTS OF PROCESS SCALABILITY

Charles Manière[a], Shirley Chan[a], Eugene A. Olevsky[a,b,*,**]

(a) Powder Technology Laboratory, San Diego State University, San Diego, USA
(b) NanoEngineering, University of California, San Diego, La Jolla, USA

**Abstract**

The microwave sintering homogeneity of large and complex shape specimens is analyzed. A new approach enabling the fabrication of complex shapes ceramics via 3D printing and microwave sintering is presented. The use of a dental microwave cavity is shown to enable a substantial level of densification of complex shape components while restricting the grain growth. The homogeneity of the processed samples during microwave sintering is studied by an electromagnetic-thermal-mechanical simulation. The realistic densification behavior, that phenomenologically takes into account the microwave effect, is included in the modeling framework. The simulation indicates the sharp correlation between the microwave field distribution in the cavity, the temperature profile, and the specimen's shape distortion.

___________________________________
* Corresponding author: **EO**: *E-mail address*: eolevsky@mail.sdsu.edu
**Fellow of the American Ceramic Society

**Nomenclature**

$\theta$ Porosity

$\dot{\theta}$ Porosity rate (s$^{-1}$)

$\underline{\sigma}$ Stress tensor (N.m$^{-2}$)

$\underline{\dot{\varepsilon}}$ Strain rate tensor (s$^{-1}$)

$\varphi$ Shear modulus

$\psi$ Bulk modulus

$Pl$ Sintering stress (Pa)

$\underline{\mathbb{1}}$ identity tensor

$\alpha$ Surface energy (J.m$^{-2}$)

$r$ Grain radius (m)

$tr(\underline{\dot{\varepsilon}})$ Trace of strain rate tensor (s$^{-1}$)

$A$ Material viscosity (Pa.s)

$A_0$ Viscosity preexponential constant (Pa.s)

$Q$ Viscosity activation energy (kJ.mol$^{-1}$)

$R$ Gas constant 8.314 (J.mol$^{-1}$.K$^{-1}$)

$\rho$ Density (kg.m$^{-3}$)

$C_p$ Heat capacity (J.kg$^{-1}$.K$^{-1}$)

$T$ Temperature (K)

$\kappa$ Thermal conductivity (W.m$^{-1}$.K$^{-1}$)

$Q_e$ Heat source (W.m$^{-3}$)

$\varphi_{rsa}$ Surface to ambient radiative heat flux (W.m$^{-2}$)

$\sigma_s$ Stefan Boltzmann constant (5.67E-8 W.m$^{-2}$K$^{-4}$)

$\epsilon$ Emissivity

$T_{air}$ Air temperature (K)

$\varphi_{csa}$ Convective heat flux (W.m$^{-2}$)

$h_{ia}$ Surface conductivity (W.m$^{-2}$.K$^{-1}$)

$J$ Surface radiosity (W.m$^{-2}$)

$G$ Irradiation flux (W.m$^{-2}$)

$n$ Refractive index

$e_b(T)$ Surface radiation produced (W.m$^{-2}$)

$\rho_r$ Reflectivity

$\alpha_a$ Absorptivity

$\varphi_{rss}$ Net inward radiative heat flux (W.m$^{-2}$)

$\mu_r$ Complex relative permeability

$\varepsilon_r$ Complex relative permittivity

$\mu_r''$ Relative permeability imaginary part

$\varepsilon_r''$ Relative permittivity imaginary part

$k_0$ The vacuum wave number (rad.m$^{-1}$)

$\sigma$ The electric conductivity (S.m$^{-1}$)

$\varepsilon_0$ The vacuum permittivity (8.854187817...×10$^{-12}$ F.m$^{-1}$)

$\mu_0$ The vacuum permeability (1.2566370614...×10$^{-6}$ T.m/A)

$j$ The complex number

$\omega$ The angular frequency (rad.Hz)

$t$ The time (s)

**E** Electric field (V.m$^{-1}$)

**H** is the magnetic field intensity (A.m$^{-1}$)

# 1. Introduction

Microwave sintering process has demonstrated a high potential for a rapid and volumetric heating[1–3] of ceramic/metal[4] samples while preserving nanometric or submicronic microstructures[5–7]. The direct interaction of microwaves with a specimen or tooling opens broad possibilities for direct/hybrid heating configurations[8] and non-conventional modes of sintering, such as microwave flash sintering, which enables the contactless sintering of ceramics[9,10] or metals[11,12] in few seconds. Microwave sintering is also a potentially low-cost technology providing a very selective and energy efficient heating.

Despite all these advantages, microwave sintering faces the challenge of control and homogeneity of the heating. The regulation of the microwave heating is difficult because it requires an advanced non-contact instrumentation and the accommodation of complex phenomena such as the resonance, the unequal microwave distributions in the cavity, and the thermal runaway of some materials[13]. The latter phenomena are also the main factors of an inherent heating instability of the microwave process that generates hot spots, densification gradients and distortions of the samples[14]. The inhomogeneities of this process can be countered via: the control of the microwave power, different microwave configurations (electric/magnetic), and the use of susceptors. Susceptors are materials which can be heated easily under microwave illumination[8]. They help the heating of low dissipative materials and, in appropriate conditions, the homogenization of the heating at elevated temperatures[15].

The establishment of stable conditions of sintering under microwave heating requires the good comprehension of different physical phenomena closely coupled to each other. The electromagnetic part governs the wave distribution which dissipates the heat in some specific area, the heat transfer governs the distribution of the heat by: conduction, convection and radiation. Finally, the temperature distribution has a direct influence on the densification of the sample governed by the continuum mechanics-based constitutive models. In consequence, a comprehensive simulation should be Multiphysics-based, and it should account for the close interaction between all the model parameters that evolve with the temperature and relative density. We previously established an electromagnetic-thermal-mechanical (EMTM) simulation framework for microwave sintering showing the importance of the coupled physics interactions in the stabilization of the microwave heating process[15].

In the present work, the EMTM simulation is used to study the homogeneity of a gear shape zirconia sample's microwave sintering. The preparation of the sample from ceramic slurry will be detailed first. Then, the sintering specifics for a large size specimen will be discussed. The simulation will be employed to understand the microwave conditions inside the cavity. Finally, the thermal homogeneity and the sample shape distortions will be discussed through the comparison of the experimental and simulation outcomes.

## 2. Experimental procedure

In this section the preparation of the green sample by a combined slurry casting and 3D printing approach is described. The microwave sintering conditions are detailed afterwards.

### 2.1. Preparation of a complex shape slurry-based ceramics using 3D printed tooling

In this work, the targeted green sample should have a 35 mm diameter gear shape. An approach similar to injection molding is employed to impose the gear shape onto a zirconia slurry. Polymer (ABS) die and punch tools are generated first by 3D printing (GETECH prusa i3) see figure 1a. Then, a zirconia slurry is prepared by adding distilled water to a zirconia powder (Tosho TZ-3YS average grain 50 nm) to obtain a viscous slurry. The slurry is cast in the die-punch assembly reported in Figure 1a and demolded by sliding the punches against the die. The gear is then taken out and dried at room temperature during 12h (see figure 1b). When dry, a 4 mm central hole is drilled. This process enables the rapid fabrication of a 43 % dense gear shape green sample without binder.

### 2.2. Microwave sintering process

The microwave sintering experiment has been carried out using a Bloomden dental microwave furnace (model PMF15B). The microwave cavity is made of a stainless steel rectangular waveguide, where a magnetron (WITOL 2M343K E625, 2.45 GHz) is connected, and a 135 mm diameter cylindrical furnace area. The heating area is thermally insulated by a fibrous alumina-silica material (80% $Al_2O_3$ - 20% $SiO_2$), two SiC susceptors are employed to heat a 50 mm diameter and 30 mm height applicator area. A zirconia balls bed is employed to prevent the adhesion of the sample to the support surface (see cavity in Figure 2). The gear sample is introduced in this area on the zirconia balls. Concerning the sintering cycle, a 1300 W microwave power is turned on during 2h20min. This step is followed by 2h of natural cooling. A pyrometer (MIKRON infrared model 140) has been adapted to record the lateral sample temperature.

## 3. Theory and calculation

An electromagnetic-thermal-mechanical (EMTM) simulation has been employed to study the microwave field distribution in the cavity, the heating and the densification of the powder. The governing equations of the pressure-less sintering and microwave heating are detailed first. The boundary conditions of the EMTM simulation are detailed afterwards.

### 3.1. Modelling of the pressure-less sintering

The sintering can be modeled via the continuum theory of sintering[16–18]. Assuming a linear viscous diffusive mechanism (usually governing the pressure-less sintering of ceramics), the main expression that relates the strain rate and stress tensors of a continuum compressible medium is:

$$\underline{\sigma} = A\left(\varphi \underline{\dot{\varepsilon}} + \left(\psi - \frac{1}{3}\varphi\right) tr(\underline{\dot{\varepsilon}})\underline{\mathbb{1}}\right) + Pl\underline{\mathbb{1}} \qquad (1)$$

with the expressions:

$$A = A_0 T exp\left(\frac{Q}{RT}\right) \qquad (2)$$

$$\varphi = (1-\theta)^2 \qquad (3)$$

$$\psi = \frac{2}{3}\frac{(1-\theta)^3}{\theta} \qquad (4).$$

The shear $\varphi$ and bulk $\psi$ moduli can be determined by a set of mechanical tests[19–21], here they are determined in accord with Ref[22]. The sintering stress expression depends on porosity, on the average particle radius $r$ and on the surface energy $\alpha$:

$$Pl = \frac{3\alpha}{r}(1-\theta)^2 \qquad (5).$$

The volume change and strain rate components are related by the mass conservation equation:

$$\frac{\dot{\theta}}{(1-\theta)} = tr(\underline{\dot{\varepsilon}}) \qquad (6).$$

The above equations are implemented in the finite element code to describe the densification of the sample. For small specimens and assuming no thermal gradients, the strain rate tensor's trace expression is $tr(\underline{\dot{\varepsilon}}) = 3\dot{\varepsilon}_r$. It is then possible to simplify the above-mentioned equations:

$$\dot{\theta} = \frac{-3\alpha(1-\theta)^3}{r\psi A_0 T exp\left(\frac{Q}{RT}\right)} \qquad (7).$$

This equation is used for the analytical modeling of densification and for the identification of the powder densification behavior (parameters $A_0$ and $Q$).

*3.2. Microwave heating*

The EMTM simulation allows to couple the microwave heating with the mechanical model previously described. The mechanical model predicts the sample densification based on the calculated temperature field. The microwave heating can be modeled using the microwave field distribution in the cavity, which is based on the Maxwell's equations[15]:

$$\nabla \times (\mu_r^{-1} \nabla \times \boldsymbol{E_r}) = k_0^2 \left(\varepsilon_r - \frac{j\sigma}{\omega \varepsilon_0}\right) \boldsymbol{E_r} \qquad (8)$$

with $\boldsymbol{E_r}$ defined by the harmonic electric field expression $\boldsymbol{E} = \boldsymbol{E_r} exp(j\omega t)$.

The heat transfer part of the EMTM model is governed by:

$$\rho C_p \frac{\partial T}{\partial t} + \nabla . (-\kappa \nabla T) = Q_e \qquad (9)$$

$$Q_e = \frac{1}{2}(\varepsilon_0 \varepsilon_r'' \boldsymbol{E}^2 + \mu_0 \mu_r'' \boldsymbol{H}^2) \qquad (10)$$

The convection of the air in the 50 mm applicator can be modelled through fluid dynamic equations[23]. However, the coupling of Maxwell's equations, heat transfer, nonlinear mechanics and fluid dynamics in a 3D model would require a significant computation power. In consequence, the conducted simulations are reduced to the EMTM coupling without the fluid dynamics part and taking into account the conduction/radiation heat transfer. This assumption is reasonable for the considered case, because the heated area is very small (50 mm diameter and 30 mm height). Under these conditions the computation time is about 10 hours (intel core i7 3.4 GHz, 32 Mo RAM).

*3.3. Boundary conditions*

The external boundary of the 135mm diameter heating area insulation-stainless-steel/air surface is subjected to a convective flux loss and to the surface-to-ambient thermal radiation described by the equations:

$$\varphi_{csa} = h_{ia}(T_{air} - T) \tag{11}$$

$$\varphi_{rsa} = \sigma_s \epsilon (T_{air}^4 - T^4) \tag{12}.$$

The internal solid/air interfaces are subjected to the mutual surface to surface thermal radiation. The air is assumed to be transparent and the solid bodies to be opaque (no radiation transmitted through the body). The relation between the thermal power radiated $e_b(T)$ and the incoming thermal irradiation G, is defined through the radiosity $J$ expression which is the total outgoing thermal radiative flux:

$$J = \rho_r G + \epsilon e_b(T) = \rho_r G + \epsilon n^2 \sigma_s T^4 \tag{13}.$$

With the ideal gray body simplification, we have the relation between the emissivity, absorptivity and reflectivity:

$$\alpha_a = \epsilon = 1 - \rho_r \tag{14}.$$

Then, we obtain the expression of the net inward radiative heat flux $\varphi_{rss}$:

$$\varphi_{rss} = \epsilon(G - e_b(T)) \tag{15}.$$

The thermal boundary conditions of the heating area are presented in Figure 3. The electromagnetic boundaries conditions encompass: perfectly reflective metallic walls surface and a TE10 port at the extremity of the rectangular waveguide where the microwave power is inlet (see Figure 2). The mechanical boundary conditions are no-penetration/no-friction conditions at the gear/zirconia-balls interface, which allow sample volume deformation with no restrictions. The materials properties are detailed in Ref[11] and reported in Table S1. The densification properties of the zirconia powder under microwave will be detailed in the results section.

## 4. Results and discussion

This section describes the experimental outcomes of the microwave sintering of a gear, the determination of the microwave sintering densification behavior and the results of the electromagnetic-thermal-mechanical finite element simulation.

*4.1. Microwave sintering results*

The photo of the gear specimen before and after the sintering cycle is reported in Figure 4. The recorded by the pyrometer temperature shows the fast heating rate of about 80 K/min in the first 20 min and the stabilization of the temperature for the next 2 hours from 1000 °C to 1080 °C. The dried slurry has a compaction of 43 % before sintering and attains 96 % of the average densification after sintering. The overall final shape reproduces the targeted gear appearance. The microstructure reported in Figure 5 is homogeneous from the center to the edge with a submicronic average grain size of about 0.5 μm. A certain level of remaining residual porosity is present probably due to the non-ideal packing of the powder in the slurry. However, knowing that no pressure was applied to prepare the green specimen, this 96 % dense and submicronic microstructure is acceptable. Some defects appear on the edge of the gear teeth and in the central hole. These defects are probably due to the slurry demolding and the manual drilling of the green specimen. Between the center and the edge some differences in height 3.78 mm ± 70 μm seem to indicate a small distortion of the shape. At this stage, it is difficult to conclude if this small distortion is due to the slurry demolding or it is originated from the intrinsic thermal gradients appearing during the microwave heating process[14,15]. To answer this question, the EMTM Multiphysics simulation of the microwave sintering process is carried out. The purpose of this simulation is to reveal the magnitude of the thermal instability in the gear during the heating and the degree of the resulting distortion of the sample.

*4.2. Sintering model*

The sintering behavior of yttria stabilized zirconia has been investigated through the sintering data of Wroe and Rowley[24] with and without microwave. They showed that under the microwave heating conditions sintering happens at a temperature about 75 K lower (in 10 K/min ramping condition) compared to the conventional sintering (see Figure 6a). The concept of the ponderomotive effect has been introduced to explain the acceleration of the sintering kinetics during the microwave sintering of ceramics. This theory considers an acceleration of the diffusive mass transport due to the action of a high electric field gradient present at the grain boundary/pore triple point[1]. The potential of this effect to accelerate the sintering has been validated using different micromechanics models[25,26]. In the present work, we consider a lower value of the preexponential constant $A_0$ in the viscosity term $A$ (decrease of the material viscosity) to model the acceleration of the sintering under microwave. Using Wroe and Rowley[24] data and equation (7) we determined the following viscosity expressions for conventional and microwave sintering. The modeled curves are reported in figure 6a.

$$A = 8.3 \, T \, exp\left(\frac{200000}{RT}\right) \quad \text{Conventional sintering} \qquad (16)$$

$$A = 4.7 \, T \, exp\left(\frac{200000}{RT}\right) \quad \text{Microwave sintering} \qquad (17)$$

Using equation (17) and the experimentally determined microwave thermal cycle (measured by the pyrometer), we obtained the densification curves for microwave and conventional densification behaviors (Figure 6b). In contrast to the conventional sintering behavior, the microwaves densification behavior looks consistent with the final density measured on the gear. In consequence, we chose the parameters of equation (17) in the EMTM simulation. It is very interesting to note that the difference between microwave and conventional sintering, which represents an improvement of only 100 K of the sintering temperatures in 10 K/min ramping conditions, introduces a higher impact in the holding regime. This indicates that the holding regime is more sensible to the accelerated sintering kinetics (identified in 10 K/min here) probably due to the longer time of sintering. In terms of the pure analytical modeling (Figure 6), this seems to explain the 96 % densification in about 2 hours of holding at 1080 °C. However, the validation of this hypothesis will ideally require microwave dilatometry data with calibrated temperatures, which is a very challenging instrumentation problem[3,13].

*4.3. Microwave distribution in the cavity*

The distribution of the electric and magnetic fields in the cavity is reported in Figure 7. In supplemental materials (Figure S1) the microwave field distribution in the cavity without the gear and the tooling materials is represented. The very different microwave field distributions in the cavity with and without materials highlight the high impact of the susceptor, the alumina tools and the sample on the electromagnetic behavior of the cavity. The traditional strategy in the microwave sintering of ceramics consists in locating the specimen in the maximum electric field area like in the waveguide. We showed previously that the dissipation term $\varepsilon_r''$ of zirconia that increases with temperature is responsible for thermal runaway phenomena that generate high thermal instabilities in the sintered samples[11,14]. In the traditional configuration, these instabilities can be more or less countered by the use of a susceptor which decreases the electric field of the sample while providing an external heating, which helps stabilizing the temperature field[8,13,15]. In the present case, the gear is located in an area where the magnetic field is high compared to the rest of the cavity. The electric field in the gear area is apparently "shielded" by the tooling (SiC, Alumina) and is twice lower than in the waveguide where it is at its maximum for all the cavity. The design of this cavity provides more stability by deliberately reducing the electric field while having two susceptors that provide an external and stable heating. Moreover, the electric filed distribution in the 50 mm diameter applicator is more stable than the non-uniform electric field distribution in the waveguide. This helps the fabrication of larger objects without using any "wave stirrer" device.

*4.4. Electromagnetic-Thermal-Mechanical simulation*

The comparison of the minimum/maximum simulated gear temperature and pyrometer measured temperature curves (reported in Figure 8a) shows them to be in a similar range. As a consequence, the densification curves obtained by the analytical model (calculated via the pyrometer curve) and the EMTM simulation (reported in Figure 8b) are similar and in good agreement with the gear final measured relative density.

The simulated electric, magnetic field modules, temperature and relative density fields are reported in Figure 9 for the cavity and sample area. Compared to the room temperature distribution reported in Figure 7, the high temperature intensity of the electric and magnetic fields in the SiC elements (Figure 9) is very low. This "cut off" phenomenon is due to the electrical conductivity of SiC, which becomes high under elevated temperatures[8]. The microwave field is then "pushed out" of the highly conductive element that gradually becomes reflective. However, the low intensity microwave field that remains in the SiC element surface (see zoom electric field image in figure 9) is sufficient to heat the material that also becomes more dissipative. A thermal equilibrium is then created, which results in the 2 hours temperature plateau profile. The overall temperature distribution shows the maximum temperature in the lower susceptor with a vertical gradient toward the upper susceptor. The lower susceptor is then, the main heating element. A vertical temperature difference of 40-50 K is observed in the sample during the holding regime (2000-8000 s) at the sintering temperatures between 1000-1080 °C. This temperature difference generates a small distortion in the sample at 5000 s. However, this distortion is reduced at the end of the sintering. Another noticeable phenomenon is the horizontal distribution of the temperature, which is lower in the center of the gear. This generates a small height difference of 0.1 mm from the center to the edge which could explain the 0.07 mm difference experimentally measured. This seems to show that the measured thickness difference is caused by the inhomogeneity of the microwave heating rather than by the slurry preparation.

## 5. Conclusions

The capability of microwave sintering to generate a 35 mm large complex shape ceramic component is investigated experimentally and by a numerical simulation. A combined experimental approach using 3D printing and slurry casting is successfully carried out to generate without pressure a gear shape green sample that can be sintered close to full density. The microstructure obtained is submicronic and very homogeneous.

Another aspect of this work is the development of an electromagnetic-thermal-mechanical simulation framework which is used to explain the microwave field distribution in the cavity, the heat transfer, the densification and the distortions of the sample. The distribution of the microwave field indicates that the used cavity enables a stable heating of larger samples up to 50 mm at 2.45 GHz. The densification behavior of the powder has been determined taking into account the different behavior of the powder material during microwave and conventional sintering. The temperature distribution in the cavity is relatively stable, a small distortion is reported in the middle of the densification process and at the end of the sintering; the evolution of the specimen's height observed experimentally is reproduced by simulation.

This case study shows that large complex shapes ceramic samples can be easily prepared by 3D printing and microwave sintering enabling high density at low temperatures (<1080 °C) with a limited grain growth.

**Acknowledgements**

The support of the US Department of Energy, Materials Sciences Division, under Award No. DE-SC0008581 is gratefully acknowledged.

**Figure captions**

Fig. 1: 3D printed polymer (ABS) set of die and punch to form the 35 mm zirconia gear shape slurry.

(W.m$^{-1}$.K$^{-1}$) the thermal conductivity, $\rho$ (kg .m$^{-3}$) the density, $\epsilon$ the emissivity, $\varepsilon'_r$, $\varepsilon''_r$, $\mu'_r$, $\mu''_r$ the permittivity and permeability real and imaginary part, respectively, and R the gas constant.

F1

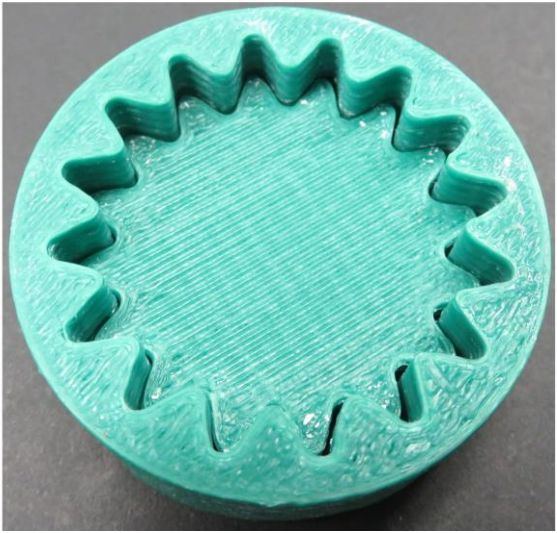
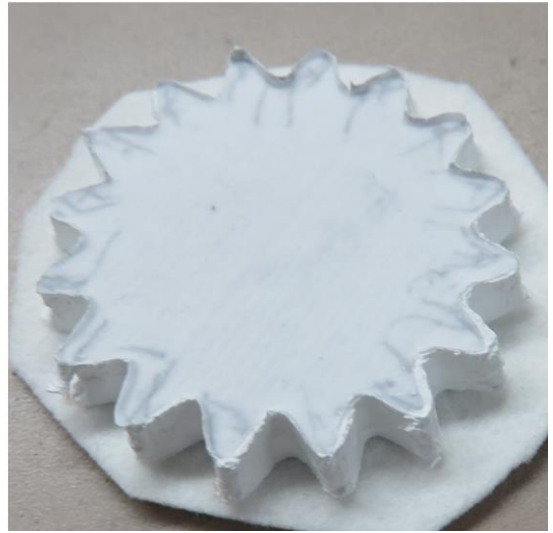

F2

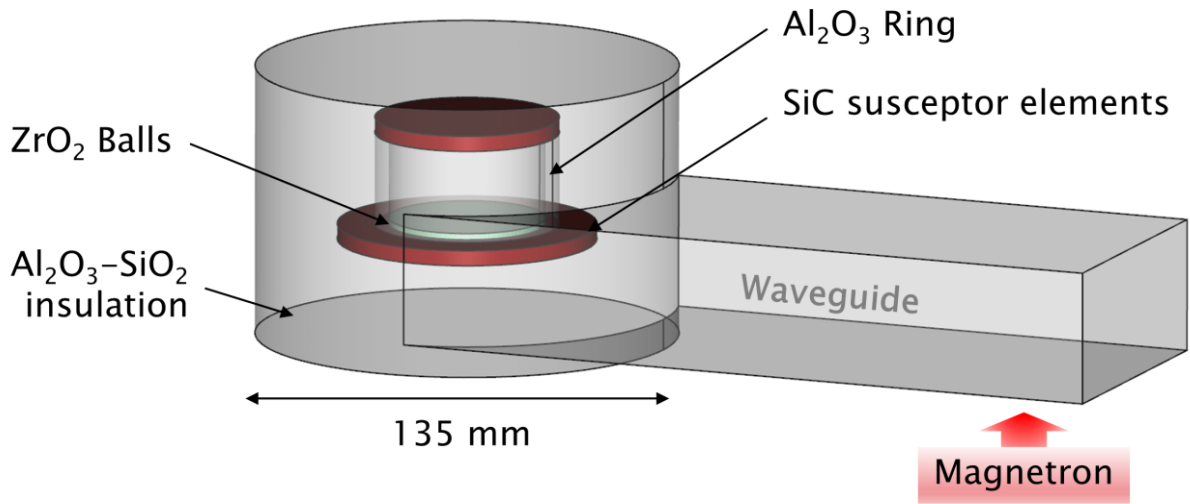

F3

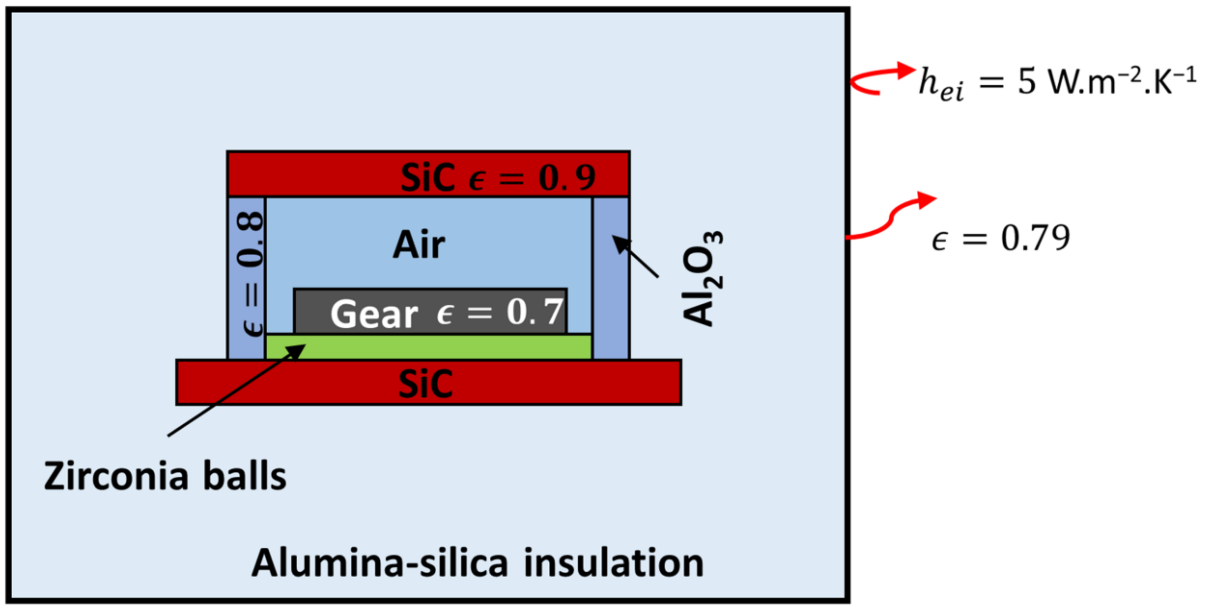

F4

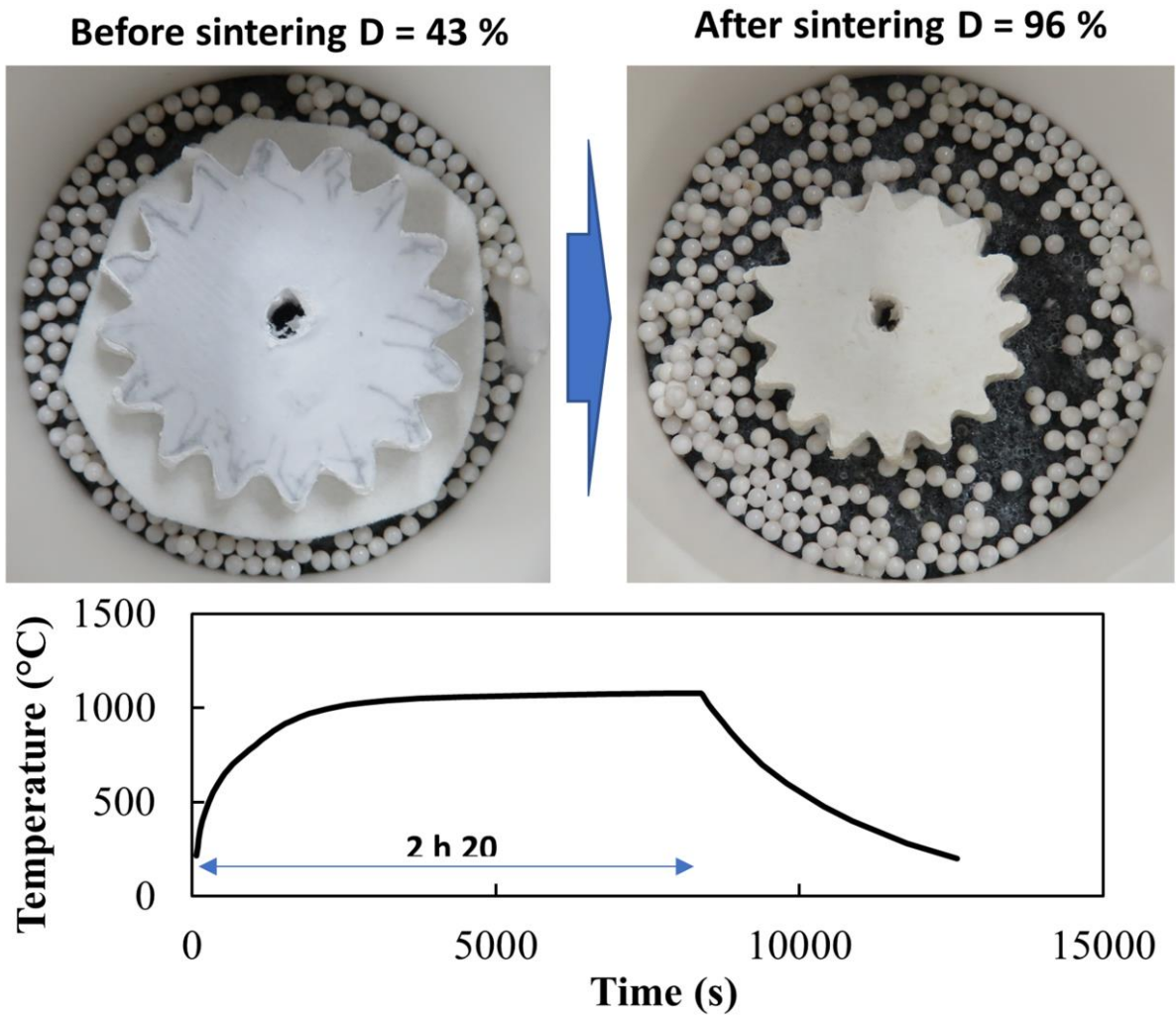

F5

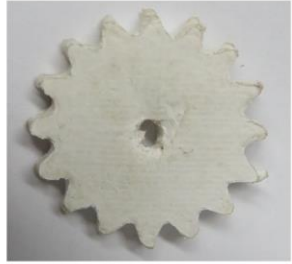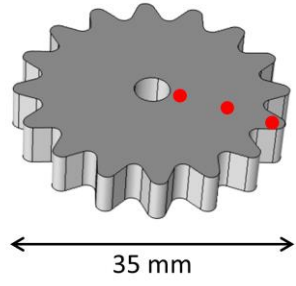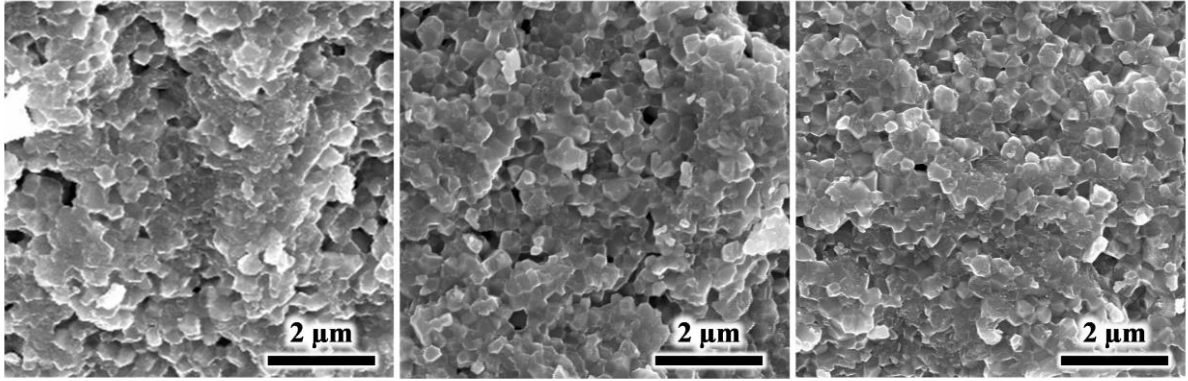

F6

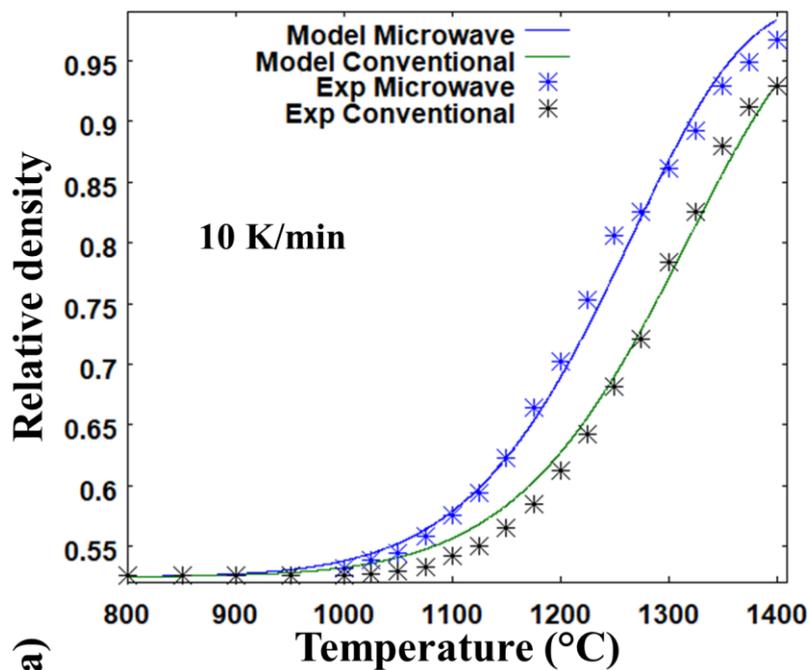

a)

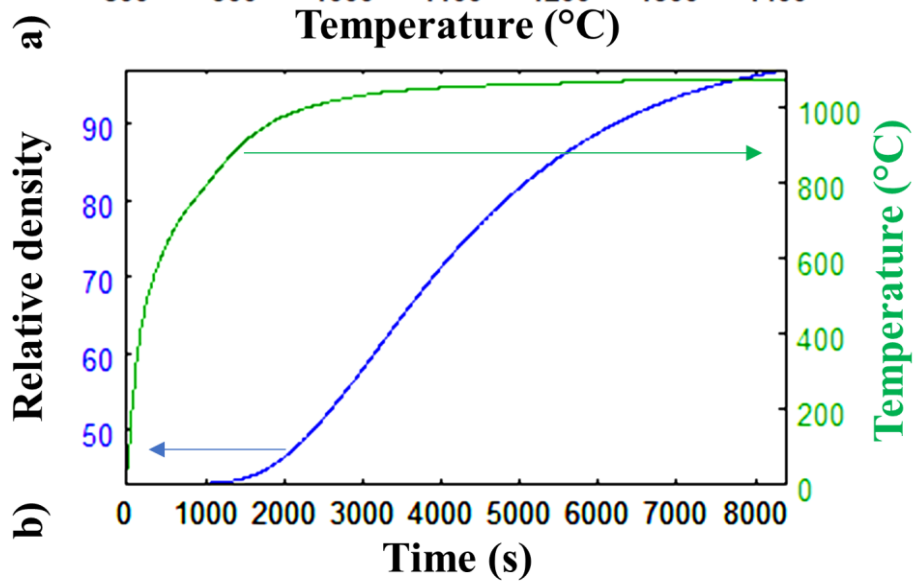

b)

F7

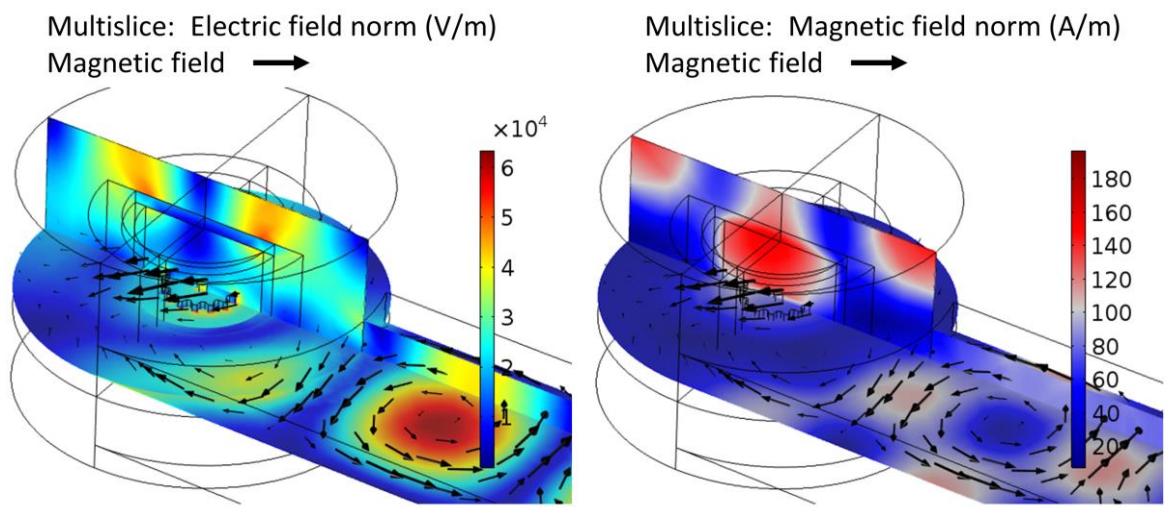

F8

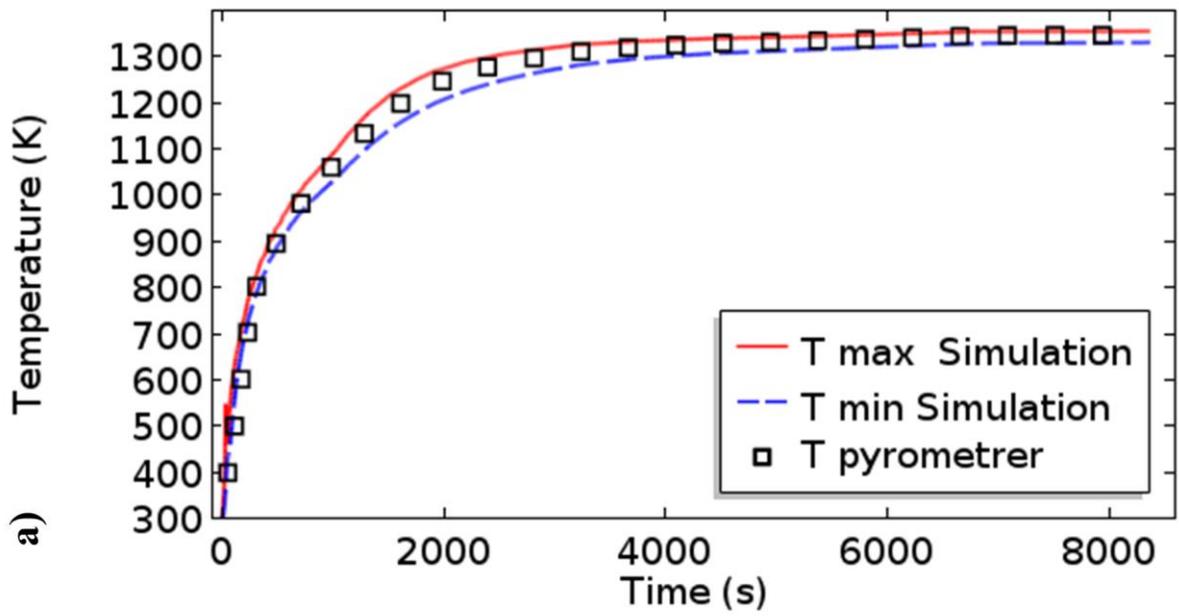

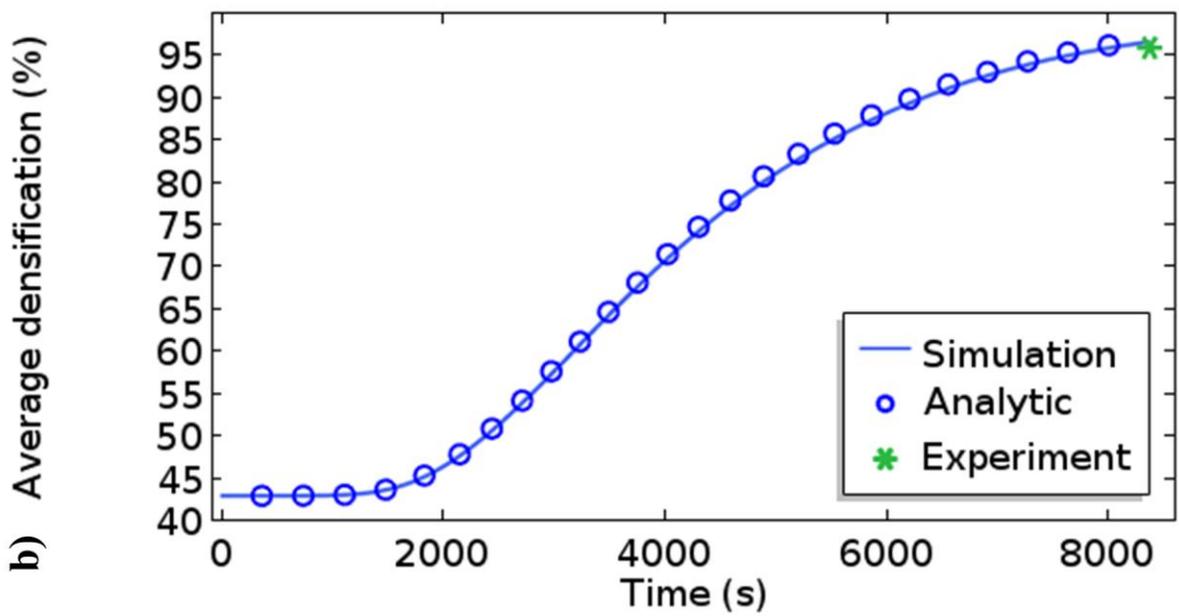

F9

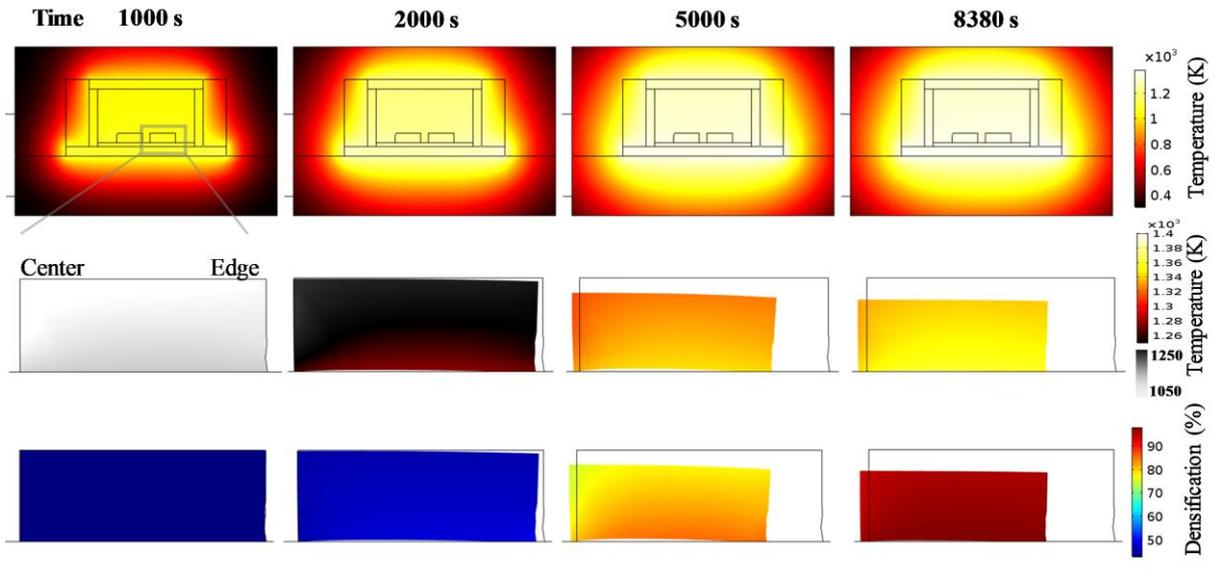